# Convergence and machine learning predictions of Monkhorst-Pack k-points and plane-wave cut-off in high-throughput DFT calculations

Kamal Choudhary, Francesca Tavazza

Materials Science and Engineering Division, National Institute of Standards and Technology, Gaithersburg, Maryland 20899, USA.

**Abstract:**

In this work, we developed an automatic convergence procedure for k-points and plane wave cut-off in density functional (DFT) calculations and applied it to more than 30000 materials. The computational framework for automatic convergence can take a user-defined input as a convergence criterion. For k-points, we converged energy per cell (EPC) to 0.001 eV/cell tolerance and compared the results with those obtained using an energy per atom (EPA) convergence criteria of 0.001 eV/atom. From the analysis of our results, we could relate k-point density and plane wave cut-off to material parameters such as density, the slope of bands, number of band-crossings, the maximum plane-wave cut-off used in pseudopotential generation, crystal systems, and the number of unique species in materials. We also identified some material species that would require more careful convergence than others. Moreover, we statistically investigated the dependence of k-points and cutoff on exchange-correlation functionals. We utilized all this data to train machine learning models to predict the k-point line density and plane-wave cut-off for generalized materials. This would provide users with a good starting point towards converged DFT calculations. The code used, and the converged data are available on the following websites: https://jarvis.nist.gov/, and https://github.com/usnistgov/jarvis .

**Introduction**

Density functional theory (DFT) is one of the most successful quantum mechanical theory based tools to characterize properties of materials[1-3]. Although density functional theory is exact in theory, its implementation requires several approximations such as the choice of basis-set, exchange-correlation[4-6] functional, mesh-size for Brillouin zone[7, 8] (BZ) integration and plane-wave cut-off[1, 9] for plane-wave basis. A systematic statistical study of how these approximating "parameters" should be chosen is still missing. In this work, we focus on two of those approximations: the plane wave cutoff and the k-point density. They are examples of approximations that can be improved in a controlled, systematic way by increasing the computational cost of the calculations. In a plane wave basis code the more plane waves we include, the better the wavefunction is modeled[1, 9], and the plane wave cut-off is the parameter that controls such truncation. Similarly, the k-point mesh is the parameter that controls the BZ integration and can play a huge role in getting accurate results, especially for metals.

The total energy of the system is the most important output of a DFT calculation, and it is obtained by numerically integrating the Hamiltonian over the Brillouin zone. The k-points are a generic way to discretize such as integral. The quality of the results heavily depends on the number of these points on the mesh-grid as well as the method generating the mesh-grid itself used in such integration[10, 11]. The number of points can be arbitrarily increased to increase the precision of calculations[12]. However, the higher the number of irreducible k-points, the higher the computational cost. Therefore, finding the optimum number of k-points to determine the total energy within a specified tolerance (i.e. "converging" on the k-point mesh) is extremely important. While most of the important physical quantities are related to energy differences (such as elastic constants and phonons), having high precision on the total energy imposes a stricter requirement than converging energy differences because high precision in energy differences might occur due to error cancellations. In principle, each property under investigation, such as bulk modulus or phonons, should be converged directly on k-points and cut-off. However, this procedure is too computationally expensive for a high throughput approach.

There are several methods to identify mesh for BZ integration, one of the most popular being Monkhorst-Pack method[7, 8]. Similarly, in principle, any wave function can be described by the superimposition of an infinite number of plane waves. However, computational constraints enforce truncation of the such an expansion i.e. all plane waves with energy above a preset threshold are ignored. Again, this truncation leads to an error in the determination of the total energy. Such an error can be reduced to be below a chosen tolerance by increasing the cutoff value ("convergence" in cutoff), however increasing the cutoff value significantly increases the computational cost. More details about the k-point and cut-off inter-dependence are given in the method section. Generally, these parameters are converged before carrying out any DFT calculations. However, recently, large DFT databases such as materials-project (MP)[13], open quantum materials database (OQMD)[14] and AFLOW[15] use fixed parameters for all the materials primarily due to their focus of quick screening. Specifically, they generally use 520 eV as plane wave-cutoff (with PAW pseudopotentials), and 1000/atom to 8000/atom k-points. Hence, it is important to understand the extent to which the choice of identical parameters for all materials is generalizable.

A recent work of Lejaeghere et al.[16] shows the reproducibility in DFT calculations across different codes, but a systematic study of the effect of plane wave parameters such as plane wave cut-off and k-points is still missing. It has been thought for years that no convergence plane wave cut-off is needed because it is determined by the elemental cut-off during PP generation. However, several recent studies showed[17-24] very high (up to 1400 eV[17-24]) plane wave-cutoff and very high k-points grid densities are needed for some materials for accurately predicting their elastic constants[24-26], phonons[12] and magnetic properties[27]. Some important previous works in elaborating the effects of convergence were done by Milman et al[28] and by Payne et al[29]. Milman et al[28] investigated the effect of finite basis correction on the equation of state properties. Payne et al[29] discussed the correlation between the k-points and plane wave-cutoff when determining total energies in the particular case of $CoSi_2$. Well-established databases such as materials project (MP), in fact, recommend increasing the cut-off and k-points for elastic constant[30] and piezoelectric calculations[31]. MP uses 7000 per-atom for elastic constant calculations. OQMD performs a set of convergence for few cases for cut-off and k-points convergence.

It is challenging to optimize all the parameter settings in DFT calculations when populating a high-throughput DFT database due to the associated computational cost. However, as a starting point, we focused on k-point density and cutoff convergence, looking for a sweet spot in computational cost and precision in this work. While automation techniques are extensively used in the field of property calculations, an automatic procedure for finding optimum parameters in the initial set-up of DFT is still missing. Here, we developed such a set-up and used it to identify converged k-point and energy cutoff parameters for at least 30000 materials. The property calculations performed with such parameters have populated the JARVIS-DFT database (https://jarvis.nist.gov ). In fact, we find that simple energy convergence results can lead to accurate energetics[32], structural[32], optoelectronic[33] and elastic[34] properties. Although in this work we focused on total energy as the property of convergence, our flexible computational framework can be used to investigate the convergence of any other property. Similarly, it can be easily modified to investigate other DFT parameters, like smearing, for instance. The computational framework is publicly available at https://github.com/usnistgov/jarvis/ .

Lastly, we used the converged k-points and energy-cutoff values to train machine learning models, so that users could predict these parameters for virtually any material with high precision before carrying out any DFT calculations. The trained machine learning models (JARVIS-ML) is also available publicly at https://www.ctcms.nist.gov/jarvisml . The training set included calculations for 30000 bulk materials available in JARVIS-DFT database. It must be noted that plane wave cut-offs are strongly correlated to the chosen pseudopotential, so our ML model predictions for cut-off values are only applicable to projector-augmented wave (PAW) calculations. The information available through the website and computational framework can be used as a fundamental tool for setting up DFT calculation before carrying out any DFT calculations.

**Method:**

The DFT calculations are performed using the Vienna Ab-initio Simulation Package (VASP)[35, 36] and the projector-augmented wave (PAW) method[5]. The list of PAW potentials used in this work is provided in the supplementary section. Please note commercial software is identified to specify procedures. Such identification does not imply recommendation by the National Institute of Standards and Technology.

In general, the DFT wavefunction is expanded in terms of a plane wave basis set:

$$\psi(r) = \sum_G c_G e^{i(G+k)r} \tag{1}$$

where, $G$ is the reciprocal wave vector, $k$ is k-point vector, $C_G$ are the expansion coefficients. There are an infinite number of allowed $G$, but the coefficient $C_G$ becomes smaller and smaller as $G^2$ becomes larger and larger. The cut-off energy $E_{cut}$ is defined[37] as: ,

$$E_{cut} = \frac{\hbar^2}{2m}|G_{cut}|^2 \qquad (2)$$

with $|G + k| < G_{cut}$

As shown in eq. 2, the $k$ and $G$ are inter-dependent, which means that the number of plane waves needed is different at each k-point. However, in our HT calculations, we decouple $k$ and $G$ during the convergence procedure due to computational cost. This way we might overestimate them, but the procedure is much faster and suitable for HT approaches. All crystal structures and their properties used in this work are available at JARVIS-DFT database. We used 0.001 eV per cell as convergence criteria for both k-points and energy cutoff. However, this value can be easily changed in the automatic procedure, because DFT calculations of some properties may require more stringent tolerances.

The k-points convergence procedure takes full advantage of the "Automatic k-mesh generation" as defined in the VASP manual. Here a "k-points line density" (L) is defined, which is related to the reciprocal lattice vectors by:

$$\vec{k} = \vec{b}_1 \frac{n_1}{N_1} + \vec{b}_2 \frac{n_2}{N_2} + \vec{b}_3 \frac{n_3}{N_3}, \quad n_1 = 0..., N_1 - 1, \quad n_2 = 0..., N_2 - 1, \quad n_3 = 0..., N_3 - 1 \qquad (3)$$

$$N_1 = \max\left(1, L \times |\vec{b}_1| + 0.5\right) \qquad (4)$$

$$N_2 = \max\left(1, L \times |\vec{b}_2| + 0.5\right) \qquad (5)$$

$$N_3 = \max\left(1, L \times |\vec{b}_3| + 0.5\right) \qquad (6)$$

Where $\vec{b}_i$ are the reciprocal lattice vectors, and $|\vec{b}_i|$ their norms.

We use the Monkhorst-Pack scheme to generate k-points, but after the generation, the grid is shifted so that one of the k-points lies on the Γ-point. We included the gamma-point because we were interested in computing quantities that require gamma-point contribution, such as optical transition for our optoelectronic database, gamma-point phonons for our elastic properties, finding multiferroic materials which have negative phonons at the gamma-point. The k-points are continuously stored in memory, to check that each of the new k-points generated by equation 3 is unique. The k-points line density starts from length 0, with Γ-point being the only k-point and is increased by 5 Å at each successive iteration if the difference between the energy computed with the new k-points and the one computed with previous k-points is more than the tolerance. After the convergence with a particular tolerance, we compute five extra points to further ensure the convergence. This procedure is repeated until convergence is found for all 5 extra points. A similar convergence procedure is carried out for the plane wave cut-off until the energy difference between successive iterations is less than the tolerance for 6 successive points. The plane wave cut-off is increased by 50 eV each time, starting from 500 eV. In both convergence procedures, we perform only single step electronic relaxation, i.e. no ionic relaxation is considered. When starting the cut-off energy convergence, we used a minimal k-point length of 10 Å. Similarly, for the k-point convergence we started with a cut-off of 500 eV. Note that complete ionic and electronic relaxation for determining converged parameters might be needed for very sophisticated calculation (such as Raman intensity calculation), but those calculations are beyond the scope of this work.

We mainly used OptB88vdW method[38, 39] for our calculations, but we also carried out about 83 local density approximation (LDA) and generalized gradient approximation with Perdew-Burke-Ernzerhof (GGA-PBE)-based[4] calculations for benchmarking purposes. The plane wave cutoff is converged using the same procedure as for the 3D bulk materials. In this work, we used Gaussian smearing (with 0.01 eV parameter) which is recommended by several DFT codes, because it is less sensitive than other methods to determine partial occupancies for each orbital. This leads to an easier DFT-SCF convergence, especially when the materials are not apriori known to be a metal or insulator, which is always the case in this work. However, it is to be emphasized that, in principle, k-points and smearing parameters should be converged together, but this requires a very computationally expensive workflow. For this reason, we choose to converge k-points and cut-off only.

In this work, we related the number of band-crossings at the Fermi level and average slope at the crossings to the number of k-points needed to achieve convergence hoping to facilitate a reasonable guess of the required k-points. The slope at each band-crossing was obtained averaging the slope of two straight lines, one connecting the energy value in the band closest to the crossing to the one just before and one connecting the energy value in the band closest to the crossing to the one just after it. Slopes at all crossing were then averaged, resulting in a single, averaged value for each metallic material. At the time of writing, the JARVIS-DFT database consists of 30039 bulk and 816 monolayer 2D materials, with formation energies, OptB88vdW (OPT) and TBmBJ (MBJ)[40] bandgaps and static dielectric constants, bulk and shear modulus and exfoliation energies for 2D layered materials. The atomic structures in the JARVIS-DFT initially were obtained from the materials-project database (MP) which again holds origin in the Inorganic Crystal Structure Database (ICSD)[41]. After obtaining the MP-PBE[4] optimized structures, we further optimized the materials with OptB88vdW method using our convergence workflow. The Optb88vdW structures give better accuracy in the lattice parameters compared to PBE[42, 43]. More details about the workflow used in developing the database are given in ref. [42, 43]. The database includes quantities such as TBmBJ gaps and dielectric functions, elastic properties, spin-orbit coupling (SOC) included bandstructures and topological spillages, solar-cell efficiency. The database contains multi-species materials up to 6 components, 201 space groups, and 7 crystal systems. Moreover, the dataset covers 1.5 % unary, 26% binary, 56 % ternary, 13 % quaternary, 2 % quinary and 1% senary compounds. The number of atoms in the simulation cell ranges from 1 to 96The k-points and the cut-off obtained from the energy convergence were used for calculating mechanical, optical and electronic properties.

**Results and discussion:**

As discussed above, the automation code takes an input structure and converges cut-off at fixed k-points and then converges k-points at fixed cut-off. An example of convergence data is shown in Fig. 1 (Bi$_2$Se$_3$, $R\bar{3}m$). Here the cut-off seems to oscillate but, as the tolerance 0.001 eV is maintained for all 5 extra steps, the cut-off energy convergence is stopped. Next, the k-points are converged in a similar way by incrementing the k-points length parameter by 5 Å each time. After the difference in energy between two consecutive k-points length parameters is smaller than the tolerance, we check 5 extra points to determine if the convergence was achieved in energy. An inset in k-points also confirms the point. The k-point length is converted into a N$_1$xN$_2$xN$_3$ ($N_i \epsilon \mathbb{Z}^+$) mesh (Eq. 4-6), which is then compared to the stored k-points list. It is to be noted that different, but similar, k-point length can give rise to identical (N$_1$, N$_2$, N$_3$), hence it is important to ensure that we do not repeat an already done calculation. Although we converge on the energy per cell (EPC) with a specific tolerance, the energy-data can be normalized with the total number of the atoms in a cell to investigate the energy per atom (EPA) convergence as well. Due to numerical reasons, as EPC values are generally greater or equal to EPA, the EPA converges equally or faster than EPC. The EPA and EPC are the same say for single-atom systems. The EPA is traditionally more widely used than EPC for k-point convergence, because traditional quantities such as formation energy per atom are 'per atom' based, hence its reasonable to converge based on EPA only. However, quantities such as energy-based stress tensor and phonons are directly derived from the energy of the simulation cell, we argue its reasonable to converge EPC though it can be more expensive than EPA. Any standard DFT code such as VASP uses the total energy convergence rather than energy per atom during a SCF procedure. Although it is generally advised to converge the physical quantity of interest say bulk modulus with respect to k-points or cut-off, we find that our EPC based calculated properties such as elastic constants[42] and dielectric[33] function are in excellent comparison with experimental or other high-level DFT results. Also, we chose EPC because we needed to find a sweet spot between computational cost and precision especially from the perspective of a high-throughput database with thousands of materials.

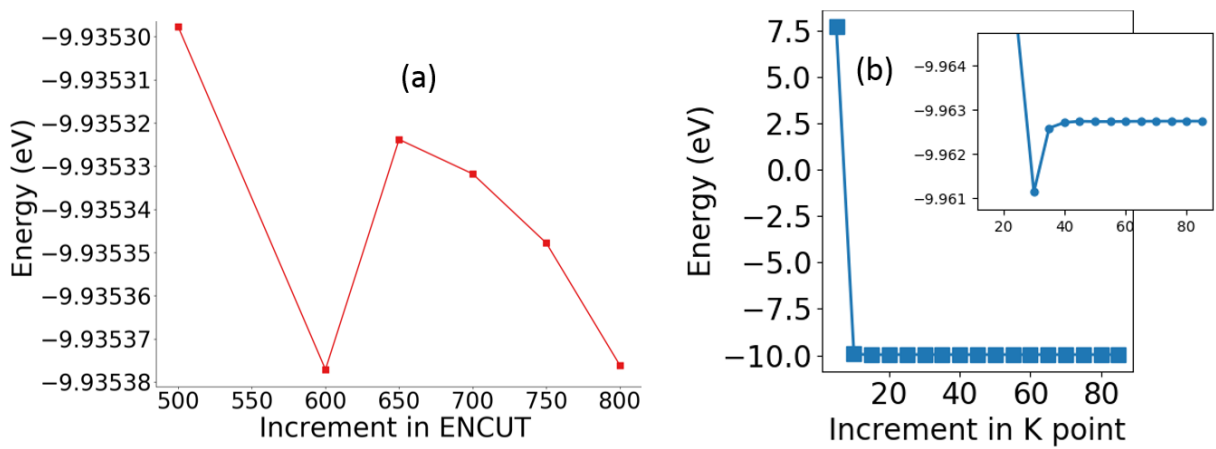

Fig. 1 An example of convergence in a) cut-off and b) k-points for $Bi_2Se_3$ (https://www.ctcms.nist.gov/~knc6/jsmol/JVASP-1067.html ).

The converged k-points and cut-off distributions for all the materials in the database are shown in Fig. 2. Fig 2a-d display data for all materials together, while Fig 2e-h (2i-l) only show results for non-metallic (metallic) systems, respectively. There are two ways of representing k-point densities: "length-based", which used the "length" as defined in Eq. 3-6 and it's in units of Å (Fig 2a), and "reciprocal atom-based", which is given by $k_{at}=(N_1 \times N_2 \times N_3)*$(number of atoms in the cell) (Fig.2b) and it's in the units of pra (per reciprocal atom). The overall behavior between the two ways of investigating k-point convergence is similar. However, the length-based k-points distribution varies from 10 Å to 200 Å (Fig 2a), while the atom-based k-points can reach very high values, up to 20000*atom, as shown in Fig. 2b. We observe similar behavior for both the EPA and EPC methods. Based on the smoothness of decay, we suggest that the length-based k-points should be preferred over per-atom formalism. This finding is in agreement with the work of Wisesa et al[10], which also provides a theoretical justification for using k-point length as a convergence parameter. Most DFT high-throughput workflows use $k_{at} \leq 8000$. For instance, MP uses $k_{at}=1000$ for most of its computed quantities, but $k_{at}=7000$ for elastic constant calculations and OQMD uses $k_{at}$ up to 7500. Comparing the EPC and EPA length-based k-point distributions (Figs. a-c-, e-g and i-k), we find that the range of the convergence values is smaller for EPA than EPC which can be explained based on a general EPC≥EPA for k-points argument. We further analyzed these distributions in terms of non-metallic (Fig. 2e-h) and metallic (Fig. 2i-l) materials. We observe that the non-metals require relatively fewer k-points than metallic compounds. The electronic band crossing at the Fermi level that only happens in metals explains why more k-points are needed in these materials to precisely integrate over the Brillouin zone.

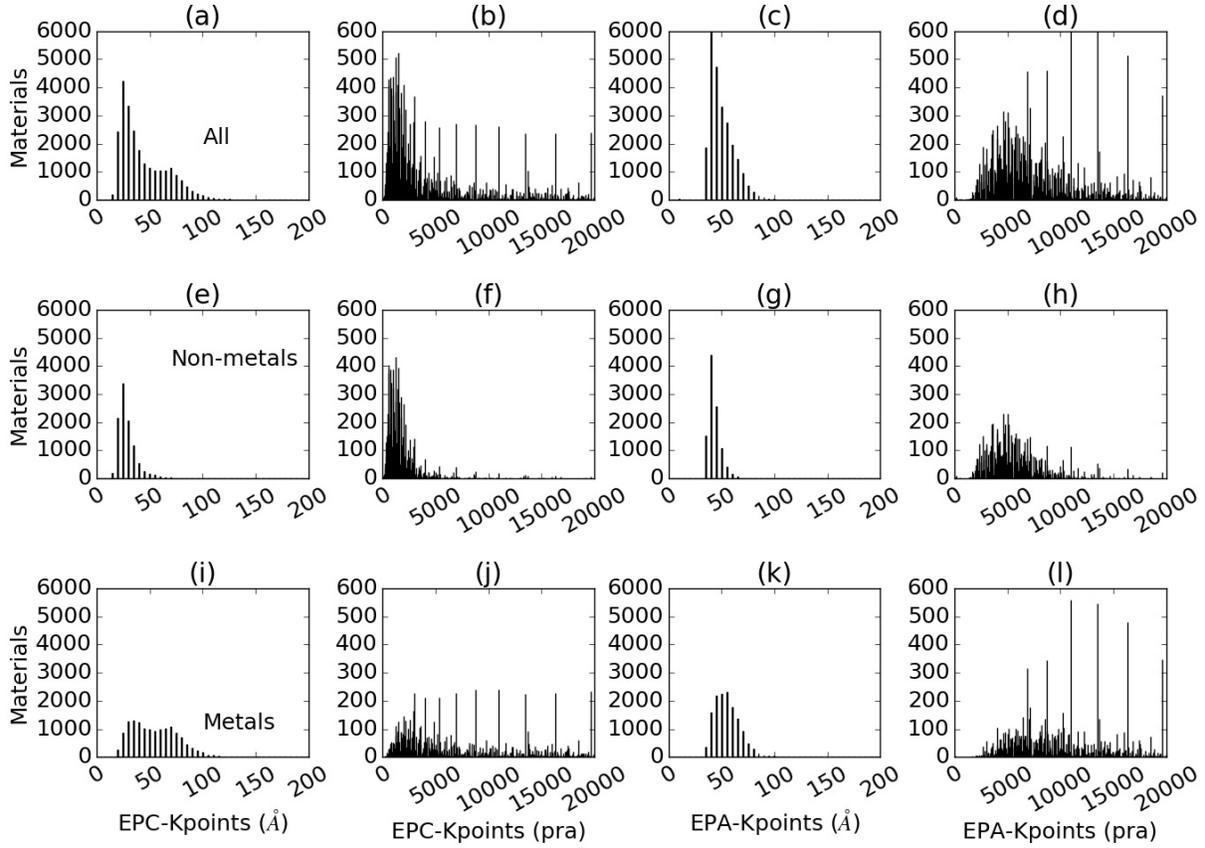

*Fig. 2 Histogram for ranges of the number of k-points. The 'energy per cell' and 'energy per atom' methods are denoted by 'EPC' and 'EPA' respectively. EPA values are derived from EPC. Fig. a) shows the length-based k-points distribution of all the materials in the database using EPC, b) the per reciprocal-atom-based distribution of k-points using EPC, c) length-based k-points distribution of all the materials in the database using EPA, d) the per reciprocal atom-based distribution of k-points using EPA. Similar distributions for non-metals are shown in e-h and for metals in i-l. Metals are in general observed to require more k-points.*

Next, the cut-off generally varies from 500 eV to 1400 eV, as shown in Fig. 3. Note we start sampling cut-off convergence from 500 eV. Therefore, the peak at 500 eV in Fig. 3, also includes materials that require cut-off less than 500 eV. This is because the goal of our work was to identify materials requiring high-cutoff. The 520 eV is usual choice of cut-off parameter in almost all the major DFT databases [13-15] . The Fig. 3b and 3c suggests that the cut-off for non-metals can be higher than the metals. The need for a larger cutoff in a non-metallic system can be explained by a general higher localization in the charge density.

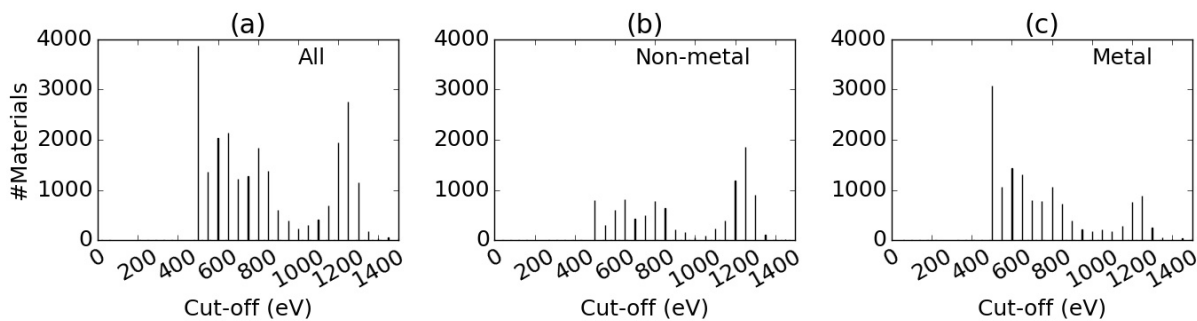

*Fig. 3 Histogram for ranges of plane-wave energy cut-off. a) for all materials, b) for non-metals and c) for metallic systems.*

Next, to gather insights into predicting converged values for k-points and cut-offs, we looked for correlations between converged values of cut-off and k-points and volumetric or basic electronic properties, such as density, volume per atom and number of electrons in a system. Our findings are displayed in Fig. 4, where we plotted the Pearson coefficient (PC)[44] for each of the examined properties. Unfortunately, none of the PC was found to be high, which means none of the examined quantities, by itself, is enough to determine what the converged values of cut-off and/or k-point length should be for a specific material. However, a combination of these quantities could be a good way to predict converged values for k-points and cut-off. The highest PC we found (0.637) was for the correlation between the maximum plane-wave cut-off used in pseudopotential generation (max_enmax) and the converged cut-off energy (Fig. 4e), which was an expected correlation. Volume per atom and density are other important quantities for cut-off and k-points, respectively. We observe that length-based k-points are more correlated to the physical quantities (as shown in Fig. 4a and 4c) than the reciprocal atom-based k-points for both EPC and EPA methods. This result also corroborates the length-based k-points to be more meaningful than the reciprocal atom-based methods as discussed earlier.

In the case of metals, we also investigated the correlation between k-points convergence and number of band crossing and average slope of the bands crossing Fermi-level. An illustrative example of band-crossing points is shown in Fig. S1. We found that the number of band-crossing is not significantly correlated to the number of k-points needed for convergence (PC=0.28 for EPC and 0.24 for EPA). However, the Pearson coefficient relating k-point length to average slope at the band crossings is among the highest PC we found, and higher than any other PC related to k-points (PC=0.62 for EPC and 0.66 for EPA), implying a relatively strong correlation.

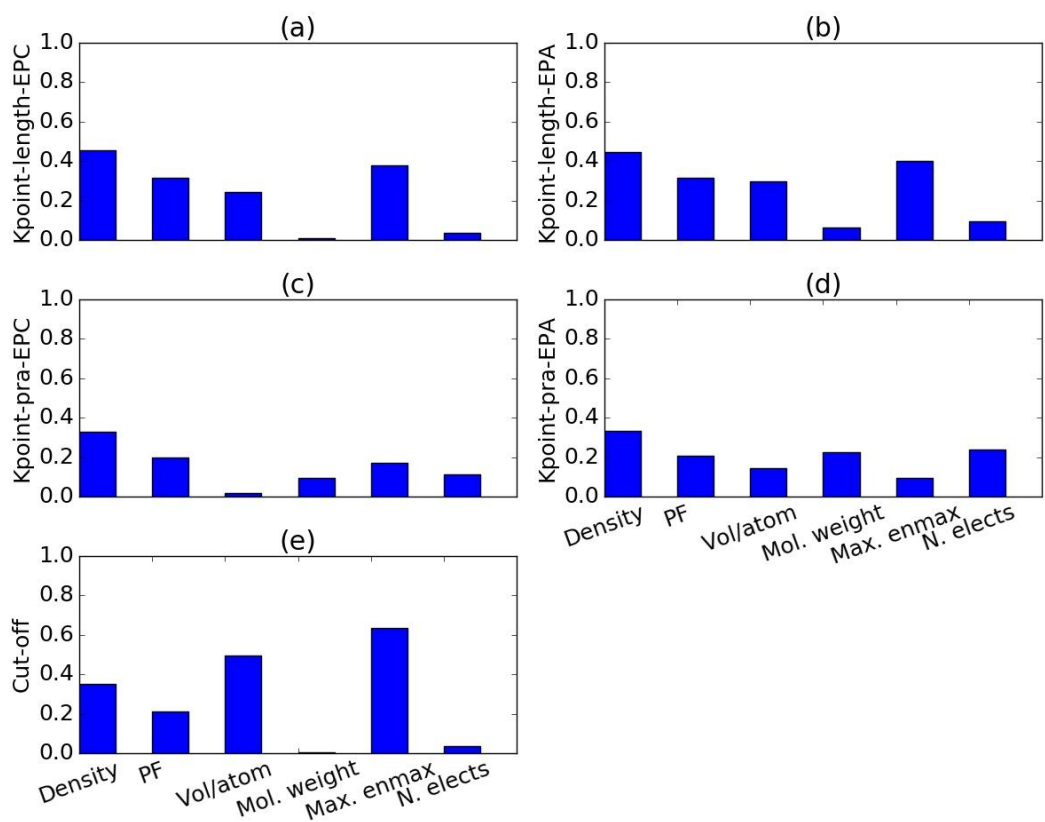

*Fig. 4 The Pearson correlation (PC) coefficients for a-d) k-points density and e) plane-wave energy cut-off with density, packing-fraction (PF), volume per atom, molecular weight, maximum elemental cut-off energy (max. enmax) and number of electrons in a system (N. elects). Fig. a and c are EPC based while b and d are EPA based coefficients. Again, a and b are based on k-point length while c and d are based on k-point densities. The best PC value, 0.66 was found for slope at the band-crossings at the Fermi-level.*

Now, we investigate the effect of crystal systems and the number of atoms per unit cell on the cut-off and k-points. In Fig. 5, we plot the average number of converged k-points and cut-off for each crystal system and the number of species in each simulation cell. As shown in Fig. 5a and 5c, cubic and hexagonal materials require higher k-points, but lower cut-off than any other system, while triclinic materials require fewer k-points but higher cutoff. Fig. 5a and 5c hence suggest an inverse relationship. Moreover, we observe that as the number of species in a system increases, the k-points decreases while the cut-off value increases, as shown in Fig. 5 b, d. In addition to the length-based distributions in Fig. 5a and 5b, we show per reciprocal atom-based k-point results in Fig. S2, which clearly indicates similar trends for both types of k-points. A possible explanation for the trend in Fig. 5b is that the more atoms are in a cell, the larger the cell is expected to be. This leads to a smaller Brillouin zone and hence a smaller k-point length is needed for convergence. The trend in Fig. 5d could be explained based on the argument that more atoms are in a cell, the chance of one of them requiring high cut-off is high. More work is obviously needed to interpret the other two figures.

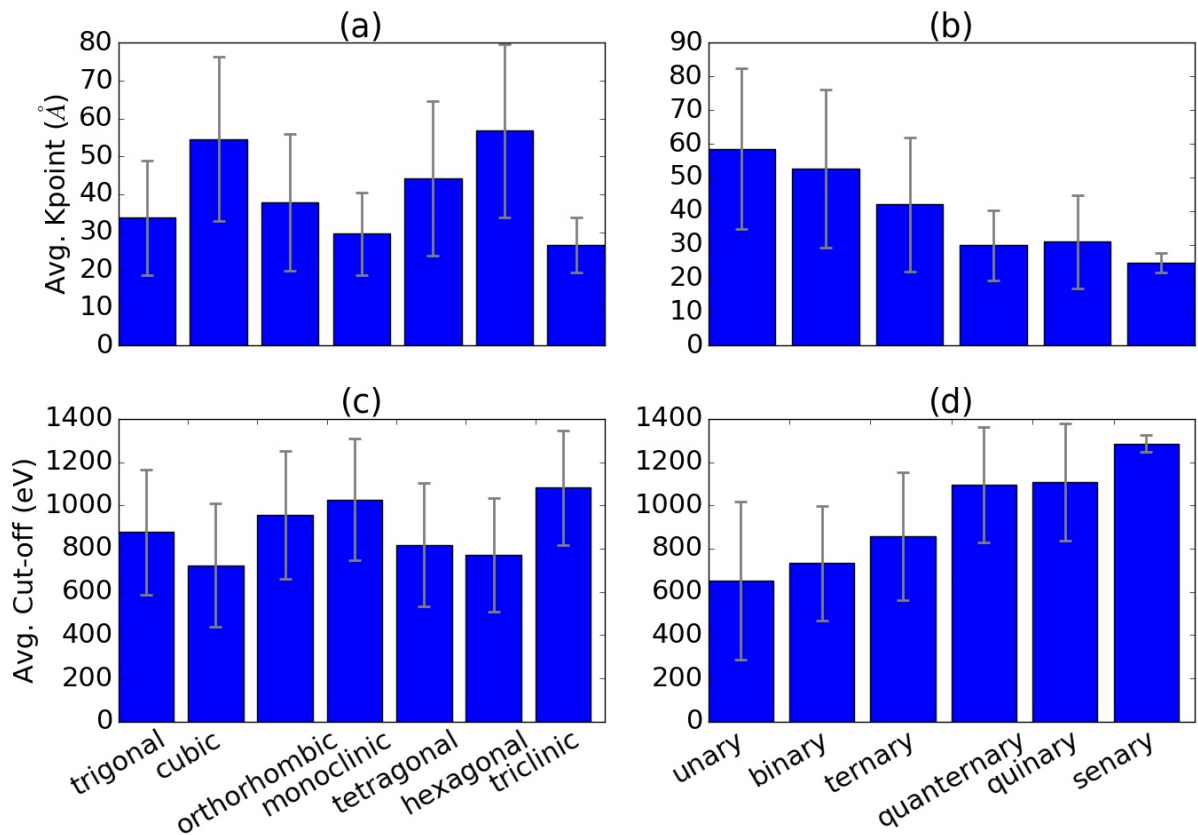

*Fig. 5 Correlation of k-points and cut-off with a, c) crystal system and b, d) number of unique elements in a simulation cell. The error bars indicate the standard deviation.*

To investigate the effect of chemical species, for each element of the periodic table we averaged the converged k-point length of all the materials in the database containing that particular element. We followed the same procedure for cutoff. The periodic table heatmap in Fig. 6 clearly shows that systems containing Be, Y, Lu, Hf, Ru, Os, Rh, Ir, Pd, Pt, Ni, Cu, B, Al, Ga and In require a higher number of k-points compared to all other elements. Similarly, a high cutoff is generally needed for systems containing Cs, La, H, Li, Na, Ba, Tc, B, C, N, O, Zn, Bi and F elements as shown in Fig. 7. Further investigation is needed to explain this behavior, which could be due to the fact that sharper features need to be described for these elements. We find that the trend in k-points length and per-atom density are similar. However, the length-based results are more uniform than the reciprocal atom-based ones, which again supports the fact that length-based formalism is a better choice for the k-point generation[10]. Interestingly, transition metals require more k-points in general, independently from being part of a metallic (Fig. 6c) or of a non-metallic (Fig. 6d) system. In terms of cut-off, highly electronegative element (such as O and F) containing compounds require a higher cut-off, as shown in Fig. 7. This is consistent with the fact that these materials generally form non-metals, which again require high cut-off as shown in Fig. 2. We emphasize that the periodic-table trends are based on all the materials (~30000) in the database rather than just elements. Note that the statistical interpretations are valid for cut-off higher than 500 eV because we start convergence from 500 eV as mentioned above also. Again, the goal of the paper was to identify materials requiring higher than usual 500 eV cut-off. While increasing the computational cost, adding more planewaves increases the precision of the calculation. Therefore, higher cut-off doesn't have any detrimental effects on the quality of the DFT data.

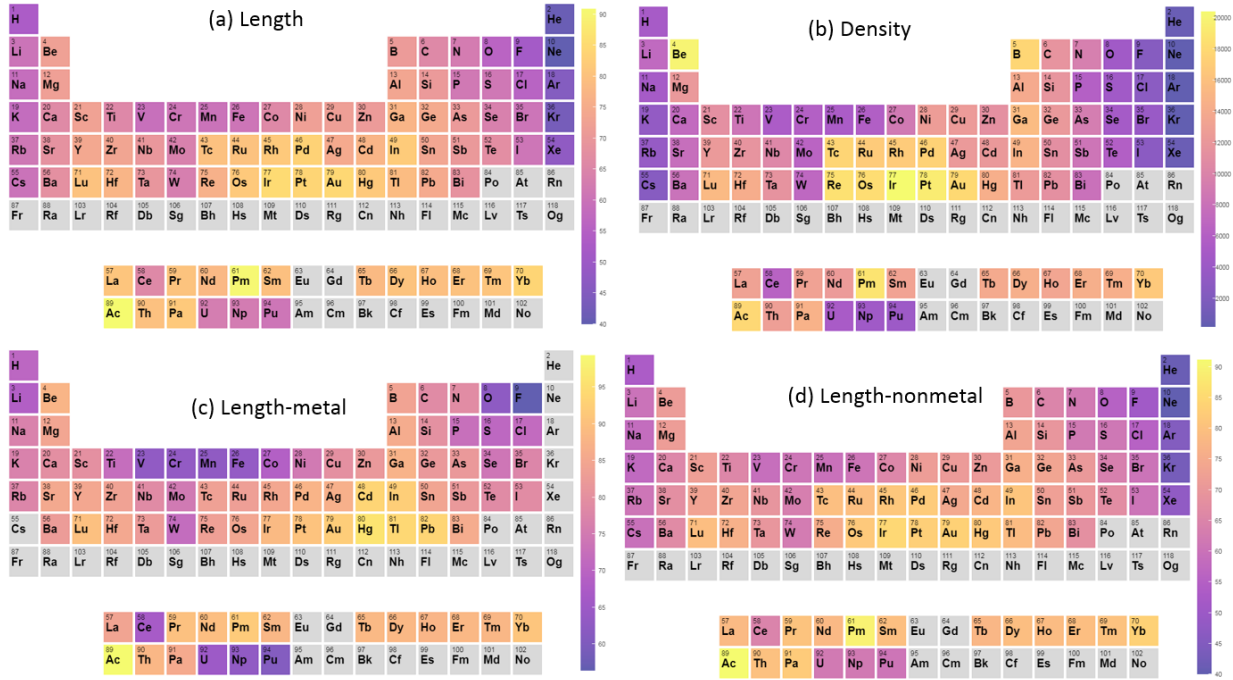

*Fig. 6 Periodic table trend for high/low k-points-requiring material constituents. The k-points of all the materials were projected on individual elements and their average contribution is shown. A) length-based k-points distribution, b) per reciprocal atom-based density distribution, c) length-based distribution for metallic materials only, d) length-based distribution for non-metallic systems only. The colorbar is in the unit of Å for length-based distributions (a, c, d) and of per reciprocal atom for density distribution (b).*

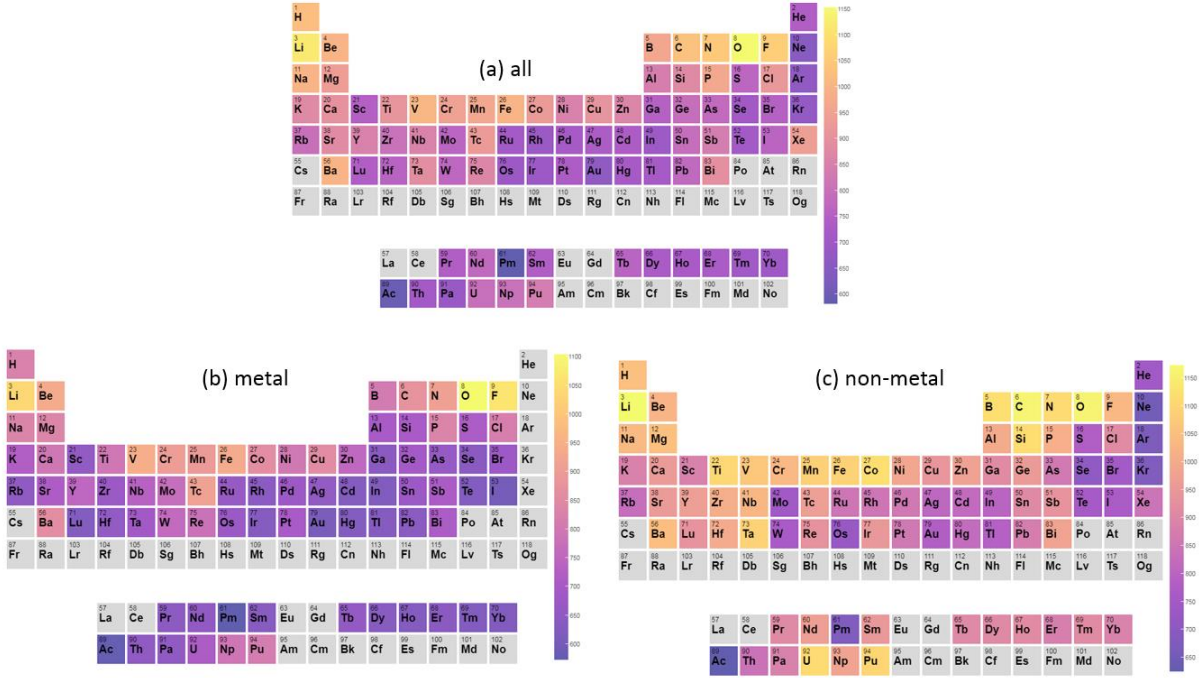

*Fig. 7 Periodic table trend for high/low cut-off-requiring material constituents. The cut-offs of all the materials were projected on individual elements and their average contribution is shown. A) cut-off distribution for all materials, b) cut-off distribution for metals, c) cut-off distribution for all non-metals. The colorbar is in the unit of eV.*

Next, we investigate if the choice of functional influences how many k-point or what cutoff is needed for convergence. We compare LDA, PBE, and OptB88vdW for 83 systems in Fig. 8a,8b. The functional comparison results show that the PBE and OptB88vdW have very similar values for convergences but LDA requires slightly fewer k-points and slightly higher cutoff. The similarity between OPT and PBE results could be attributed to the fact that both functionals utilize a density gradient, unlike LDA. However, overall the differences between convergence needs are very small among all three, hence convergence from one functional should be transferable to other functionals.

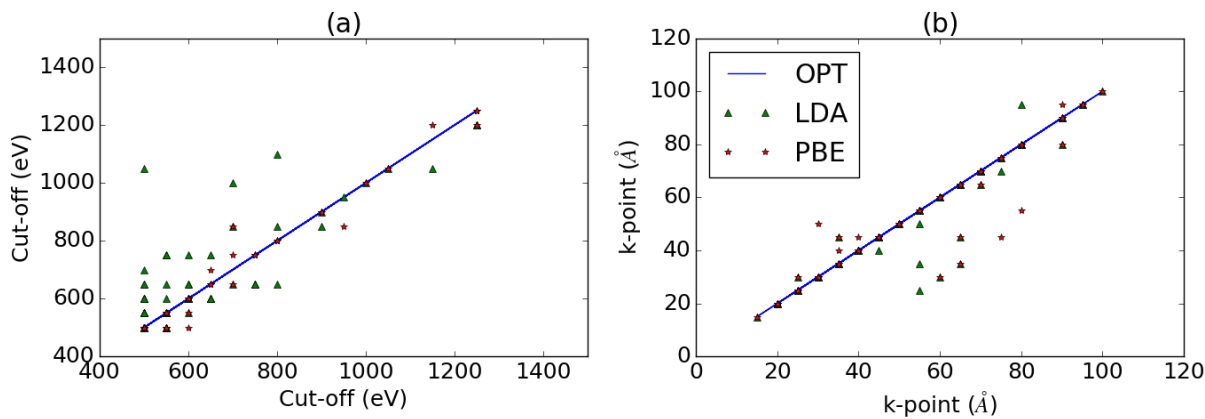

*Fig. 8 Comparison of k-points and plane-wave cut-off for different exchange-correlation (XC). In Fig. a, the cut-off is compared for LDA, PBE and OptB88vdW for 83 materials. Similarly, in Fig. b k-points in length-based representations are compared for the XCs.*

As no obvious trends were found for predicting converged cutoff and k-points during correlation study, we chose to use machine learning (ML) to predict these quantities. We train machine learning models for cut-off and k-points using Classical Force-field Inspired Descriptors (CFID)[45] and gradient boosting decision tree (GBDT). The complete dataset is available at *https://figshare.com/articles/JARVIS-ML-CFID-descriptors_and_material_properties/6870101* . Different steps involved during the ML training are described in the Jupyter notebook: https://github.com/usnistgov/jarvis/blob/master/jarvis/db/static/jarvis_ml-train.ipynb. More details about the ML training and validation are given in Ref [45]. The dataset was split into 90% and 10% for training and testing purposes. After training the ML models, the mean absolute errors (MAE) in cut-off and k-points were 85 eV and 9.09 Å for the 10 % held set. Compared to the spread of the distribution for the k-points and cut-off (Fig 2), these results promise reasonable ML predictions. This also implies the CFID are suitable descriptors for k-points and cut-off data. A web-app for predicting cut-off and k-points for a structure is available at https://www.ctcms.nist.gov/jarvisml. Since we use GBDT during the ML training, the feature importance can be obtained for the ML models. The feature-importance plot for these models is shown in Fig. 9a and 9c. Clearly, the chemical and radial distribution function (RDF) are the most important parameters during ML training. The cell-based feature such as density and packing fraction descriptors are revealed as important features as also discussed in Pearson coefficient results (Fig. 4). Some of the important chemical features were: average ionic radii, the average of ratios of melting point and heat of fusion, and electron affinity of the constituent elements. The ML performance on 10% held data is shown in Fig. 9b and 9d. The test set consisted of 998 non-metals and 1239 metallic systems. The predictions are discrete because the k-point and cut-off are increased in a step-wise manner during the actual DFT convergence.

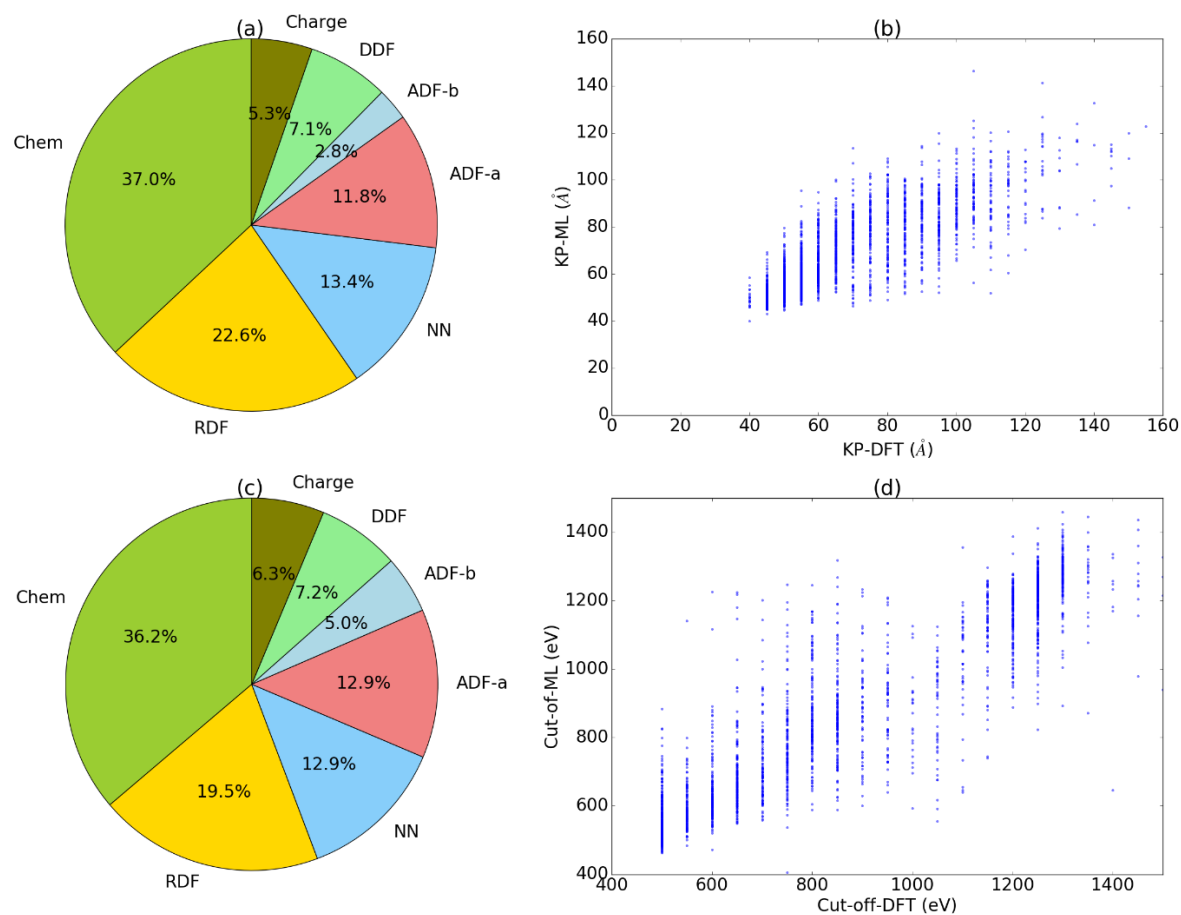

*Fig. 9 Machine learning prediction results for k-points and cut-off. a, c) feature importance plots for k-points and cut-off, b-d) prediction on 10% ML-training held set for k-points and cut-off. Here 'Chem' represents chemical descriptors, 'RDF' radial distribution function descriptors, 'NN' nearest neighbor descriptors, 'ADF' angular distribution function upto first (a) and second (b) neighbors, 'DDF' the dihedral distribution function, and 'Charge' the charge distribution descriptors.*

**Conclusions:**

In this work, we investigated how to predict the plane-wave cutoff and the number of k-points needed to reach a 0.001 eV/cell convergence in energy. Hence, we developed an automatic convergence procedure for k-points and cut-off and applied it to more 30000 materials. The computational framework is very flexible to arbitrary convergence criteria and it is publicly available at https://github.com/usnistgov/jarvis . We determined the range of the k-points and cut-off distributions. Based on the comparison between two k-point representations, we suggest that the length-based k-points should be preferred over per-atom formalism. While we determined a mean of 44 Å k-points-length and 856 eV cut-off, we recommend converging these parameters for individual materials as large outliers that we found. We demonstrated relationships between k-points, cut-off parameters and material parameters, such as density, number of electrons, the maximum cut-off used in pseudopotential generation and crystal systems. We found that cut-off used during the generation of the pseudopotential and volume per atom are highly correlated to the plane wave parameters. We also identified some of the material species that would require more careful convergence than others. We showed that the cut-off and k-points should be independent of the functional but mainly dependent on the type of materials. Although energy is chosen as the convergence property, any other property can also be converged with our computational set-up. We then developed a machine-learning model to predict the line density and cut-off for materials on-fly. We believe our results can be used as a guide before carrying out DFT calculations.

**Data availability: https://doi.org/10.6084/m9.figshare.6870101.v1** .

# Supplementary information: Convergence and machine learning predictions of Monkhorst-Pack k-points and plane-wave cut-off in high-throughput DFT calculations

Kamal Choudhary, Francesca Tavazza

Materials Science and Engineering Division, National Institute of Standards and Technology, Gaithersburg, Maryland 20899, USA

Table S1: PAW potentials used during the generation of DFT database.

| Element | VASP PAW-potential used |
|---------|------------------------|
| Ac | Ac |
| Ag | Ag |
| Al | Al |
| Ar | Ar |
| As | As |
| Au | Au |
| B | B |
| Ba | Ba_sv |
| Be | Be_sv |
| Bi | Bi |
| Br | Br |
| C | C |
| Ca | Ca_sv |
| Cd | Cd |
| Ce | Ce |
| Cl | Cl |
| Co | Co |
| Cr | Cr_pv |
| Cs | Cs_sv |
| Cu | Cu_pv |
| Dy | Dy_3 |
| Er | Er_3 |
| Eu | Eu |
| F | F |
| Fe | Fe_pv |
| Ga | Ga_d |
| Gd | Gd |
| Ge | Ge_d |
| H | H |
| He | He |
| Hf | Hf_pv |
| Hg | Hg |
| Ho | Ho_3 |
| I | I |
| In | In_d |
| Ir | Ir |
| K | K_sv |
| Kr | Kr |
| La | La |
| Li | Li_sv |
| Lu | Lu_3 |
| Mg | Mg_pv |
| Mn | Mn_pv |
| Mo | Mo_pv |
| N | N |
| Na | Na_pv |
| Nb | Nb_pv |
| Nd | Nd_3 |
| Ne | Ne |
| Ni | Ni_pv |
| Np | Np |
| O | O |
| Os | Os_pv |
| P | P |
| Pa | Pa |
| Pb | Pb_d |
| Pd | Pd |
| Pm | Pm_3 |
| Pr | Pr_3 |
| Pt | Pt |
| Pu | Pu |
| Rb | Rb_sv |
| Re | Re_pv |
| Rh | Rh_pv |
| Ru | Ru_pv |
| S | S |
| Sb | Sb |
| Sc | Sc_sv |
| Se | Se |
| Si | Si |
| Sm | Sm_3 |
| Sn | Sn_d |
| Sr | Sr_sv |
| Ta | Ta_pv |
| Tb | Tb_3 |
| Tc | Tc_pv |
| Te | Te |
| Th | Th |
| Ti | Ti_pv |
| Tl | Tl_d |
| Tm | Tm_3 |
| U | U |
| V | V_pv |
| W | W_pv |
| Xe | Xe |
| Y | Y_sv |
| Yb | Yb_2 |
| Zn | Zn |

| Zr | Zr_sv |
|----|-------|
|    |       |

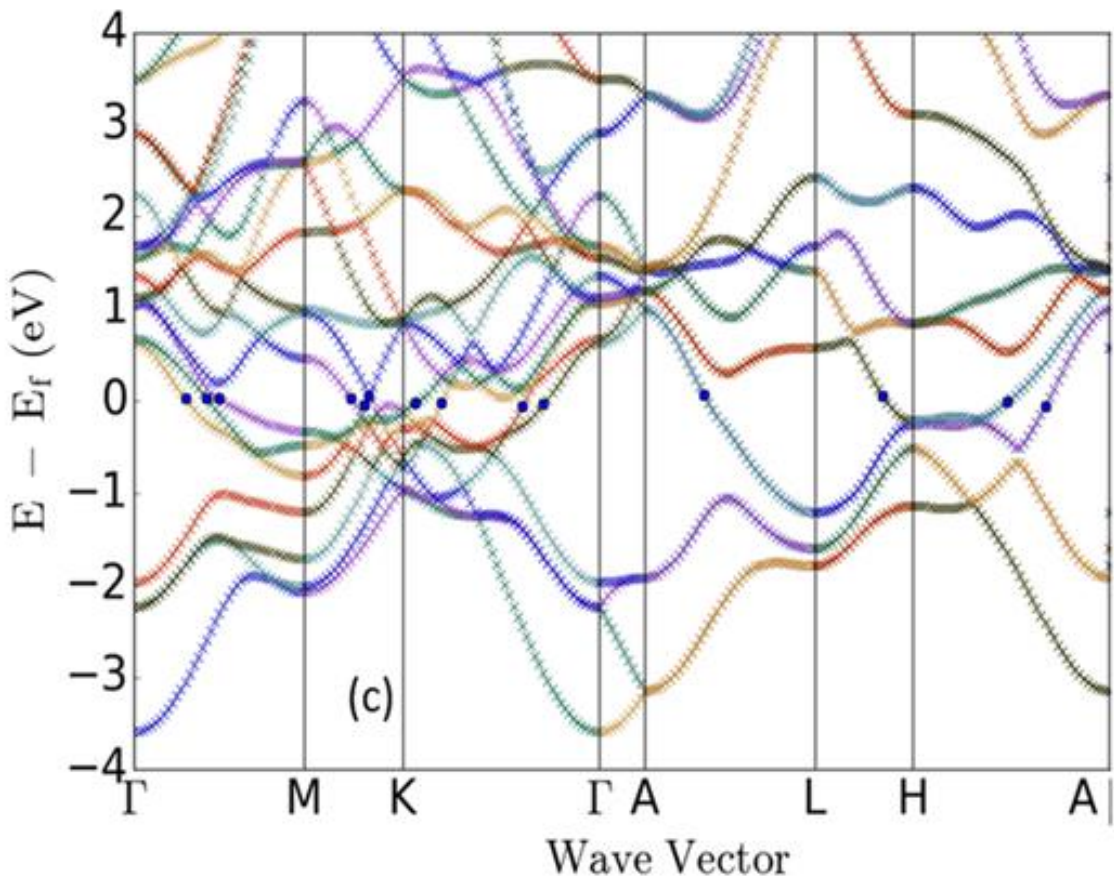

Fig. S1 An example is shown for band-crossings in Pr (JVASP-14689). The dots represent the bands crossing the Fermi-level.

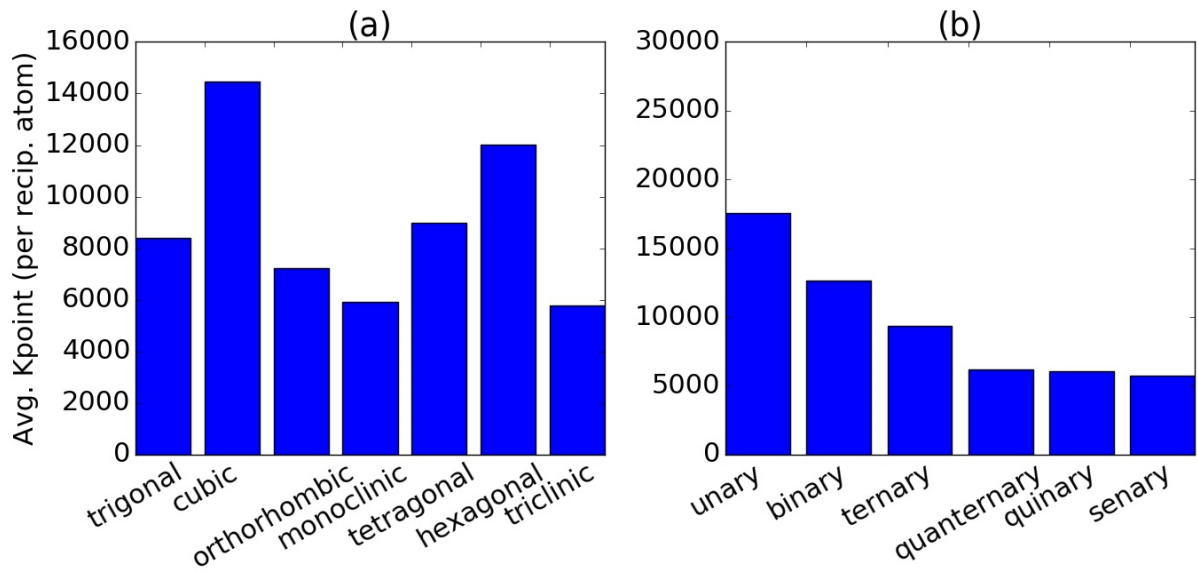

Fig. S2 Correlation of density based average k-points for the seven crystal systems and number of atoms in the cell.